\documentclass[]{aa}

\def\rmit#1{{\it #1}}              
\def\specchar#1{{\sc #1}}
\def\FeI{\mbox{Fe\,\specchar{i}}}

\def\SiI{\mbox{Si\,\specchar{i}}}
\def\CaI{\mbox{Ca\,\specchar{i}}}

\def\CaIIH{\mbox{Ca\,\specchar{ii}\,\,H}}       

\def\eg{\rmit{e.g.}}

\def\arcsec{\hbox{$^{\prime\prime}$}}

\usepackage{booktabs}
\usepackage{graphicx}
\usepackage{multirow}
\usepackage{color}
\usepackage[flushleft]{threeparttable}

\usepackage{array}
\newcolumntype{?}{@{\vrule width 2pt}}

\usepackage{hyperref}
\hypersetup{
    final=true,
    pageanchor=true,
    colorlinks=true,
    breaklinks=true,
    linkcolor=blue,
    citecolor=blue,
    urlcolor=blue,
    pdfpagemode=UseNone,
    pdftitle={},
    pdfauthor={T. Felipe et al.},
    pdfsubject={Solar Physics},
    pdfkeywords={Methods: observational -- Sun: photosphere -- Sun: oscillations  -- sunspots -- Techniques: spectroscopy-- Techniques: polarimetric}}

\titlerunning{Spiral-shaped wavefronts}   

\begin{document}



\title{Spiral-shaped wavefronts in a sunspot umbra}

\author{T. Felipe\inst{\ref{inst1},\ref{inst2}}
\and C. Kuckein\inst{\ref{inst3}}
\and E. Khomenko\inst{\ref{inst1},\ref{inst2}}
\and I. Thaler\inst{\ref{inst5}}
}


\institute{Instituto de Astrof\'{\i}sica de Canarias, 38205, C/ V\'{\i}a L{\'a}ctea, s/n, La Laguna, Tenerife, Spain\label{inst1}
\and 
Departamento de Astrof\'{\i}sica, Universidad de La Laguna, 38205, La Laguna, Tenerife, Spain\label{inst2} 
\and 
Leibniz-Institut f{\"u}r Astrophysik Potsdam (AIP), An der Sternwarte 16, 14482 Potsdam, Germany\label{inst3} 
\and
Racah Institute of Physics, The Hebrew University of Jerusalem, 91904 Jerusalem, Israel\label{inst5}
}

\abstract
{Solar active regions show a wide variety of oscillatory phenomena. The presence of the magnetic field leads to the appearance of several wave modes, whose behavior is determined by the sunspot thermal and magnetic structure.}
{We aim to study the relation between the umbral and penumbral waves observed at the high photosphere and the magnetic field topology of the sunspot.}
{Observations of the sunspot in active region NOAA 12662 obtained with the GREGOR telescope (Observatorio del Teide, Tenerife, Spain) were acquired on 2017 June 17. The data set includes a temporal series in the \FeI\ 5435 \AA\ line obtained with the imaging spectrograph GREGOR Fabry-P\'erot Interferometer (GFPI) and a spectropolarimetric raster map acquired with the GREGOR Infrared Spectrograph (GRIS) in the 10830 \AA\ spectral region. The Doppler velocity deduced from the restored \FeI\ 5435 \AA\ line has been determined, and the magnetic field vector of the sunspot has been inferred from spectropolarimetric inversions of the \CaI\ 10839 \AA\ and the \SiI\ 10827 \AA\ lines.}
{A two-armed spiral wavefront has been identified in the evolution of the two-dimensional velocity maps from the \FeI\ 5435 \AA\ line. The wavefronts initially move counterclockwise in the interior of the umbra, and develop into radially outward propagating running penumbral waves when they reach the umbra-penumbra boundary. The horizontal propagation of the wavefronts approximately follows the direction of the magnetic field, which shows changes in the magnetic twist with height and horizontal position.}
{The spiral wavefronts are interpreted as the visual pattern of slow magnetoacoustic waves which propagate upward along magnetic field lines. Their apparent horizontal propagation is due to their sequential arrival to different horizontal positions at the formation height of the \FeI\ 5435 \AA\ line, as given by the inclination and orientation of the magnetic field.}


\maketitle


\section{Introduction}

Sunspot waves are one of the fundamental dynamic phenomena found in solar magnetic structures, and have been thoroughly studied by means of observational, analytical, and numerical works over the last decades. Observational works have reported a plethora of oscillatory behaviors in sunspots. The first detection was performed by \citet{Beckers+Tallant1969}, who measured sudden periodical brightenings in the core of \CaIIH\ and K lines that they called umbral flashes. Subsequent works have reported the presence of 5 minute oscillations in the umbral photosphere \citep{Bhatnagar+etal1972,Soltau+etal1976}, 3 minute oscillations in the chromosphere \citep{Beckers+Schultz1972}, a pattern of radially outwards propagating wavefronts across the sunspot penumbra at chromospheric layers known as running penumbral waves \citep{Giovanelli1972, Zirin+Stein1972}, and also the photospheric counterpart of the running penumbral waves \citep{Musman+etal1976}, among other wave phenomena. See \citet{Thomas1985, Lites1992,Bogdan+Judge2006,Khomenko+Collados2015} for a review.

Despite the increasing literature, a complete understanding of oscillations in sunspots remains elusive. It is well established that near-surface convective motions excite p-modes \citep[\eg,][]{Goldreich+Kumar1990, Balmforth1992} in the quiet Sun. As the photospheric sunspot wave spectrum is similar to that found in the quiet-Sun regions \citep{Penn+Labonte1993}, it is expected that 5 minute umbral photospheric oscillations are driven by external p-modes. This scenario is supported by the measurements of wave absorption produced by sunspots \citep{Braun+etal1987,Braun1995}. Since the p-mode power of waves traveling towards sunspots is higher than that of waves traveling outward, it is considered that sunspot waves can be generated through mode conversion \citep{Cally+Bogdan1993} from external p-modes. Convection can also take place even in the presence of strong magnetic fields in sunspots, as shown by numerical models of magnetoconvection \citep[\eg][]{Schussler+Vogler2006} and observations \citep[\eg][]{Bharti+etal2007b}. Thus, sunspot waves can also be excited by convective motions below the umbral surface. Recently, the excitation of sunspot waves by sources located at depths up to 5 Mm beneath the surface has been claimed based on the detection of fast-moving wavefronts propagating radially outward across the sunspot photosphere \citep{Zhao+etal2015, Felipe+Khomenko2017}.   

The precise nature of the 3 minute oscillations has been a matter of debate for several decades, and a very active field of research in the 80s and 90s. Those studies tried to interpret waves with those frequencies as the result of resonances in the sunspot \citep[\eg][]{Bogdan+Cally1997}. In this theoretical picture, the resonant cavity is produced by wave reflections due to the vertical stratification of the sunspot atmosphere. Several kinds of resonant cavities have been proposed. In the scenario of photospheric fast-mode resonances \citep{Scheuer+Thomas1981}, the lower turning point of the waves is produced by sound speed gradients, and the reflection at the upper turning point is due to gradients in the Alfv\'en speed. On the contrary, the chromospheric resonance model \citep{Zhugzhda+Locans1981, Zhugzhda+etal1983,Gurman+Leibacher1984} considers that high-frequency slow magnetoacoustic waves are partially trapped between the transition region and the photospheric temperature gradients. This model has been recently refined by analyzing the multilayer structure of the sunspot atmosphere \citep{Zhugzhda2008} and by performing numerical modeling \citep{Botha+etal2011}. \citet{Fleck+Schmitz1991} proposed that long period disturbances can excite oscillations at the cutoff frequency, providing an alternative explanation to the 3 minute waves with no need for a chromospheric cavity.

Nowadays, the most favored model to explain the chromospheric 3 minute oscillations is the propagation of slow magnetoacoustic waves from the photosphere to higher atmospheric layers \citep{Centeno+etal2006}. The amplitude of the high frequency waves (with frequency above the cutoff value, like waves in the 3 minute band) increases with height due to the drop of the density, and they dominate over the evanescent 5 minute oscillations. This model has been used to explain the chromospheric running penumbral waves \citep{Bloomfield+etal2007b,Madsen+etal2015}, to determine the formation height of several spectral lines \citep{Felipe+etal2010b}, to estimate the energy contribution of those waves to the chromospheric heating \citep{Felipe+etal2011, Kanoh+etal2016}, and to reconstruct the magnetic field inclination from measurements of the cutoff frequency \citep{Yuan+etal2014, LohnerBottcher+etal2016}, among other studies. The source of these waves can be located inside the sunspot, through magnetoconvection occurring in light bridges and umbral dots \citep{Chae+etal2017}, or they can be externally driven by p-modes \citep{KrishnaPrasad+etal2015}.   

The propagation of slow magnetoacoustic waves is guided by magnetic field lines, and only waves with frequency above the cutoff frequency can propagate. In the presence of inclined magnetic field, the value of the cutoff frequency is lower due to the reduced gravity in the direction of wave propagation \citep{Bel+Leroy1977, Jefferies+etal2006}. This fact suggests that the pattern of the observed wavefronts must be determined by the magnetic field topology and the thermal structure of the sunspot. Interestingly, some recent works have detected the presence of umbral wavefronts showing evolving spirals \citep{Sych+Nakariakov2014, Su+etal2016}. In this paper, we identify a new spiral-shaped wavefront event, and we explore how it is related with the geometry of the magnetic field extracted from spectropolarimetric inversions. In Section \ref{sect:observations} we describe the observations and the analysis methodology. Section \ref{sect:results} illustrates the results, including the description of the spiral and its evolution, together with the inferred magnetic field topology. Finally, in Section \ref{sect:conclusions} we discuss and summarize the main findings of this work.

\section{Sunspot observations and data analysis}
\label{sect:observations}

\begin{figure}[!ht] 
 \centering
 \includegraphics[width=9cm]{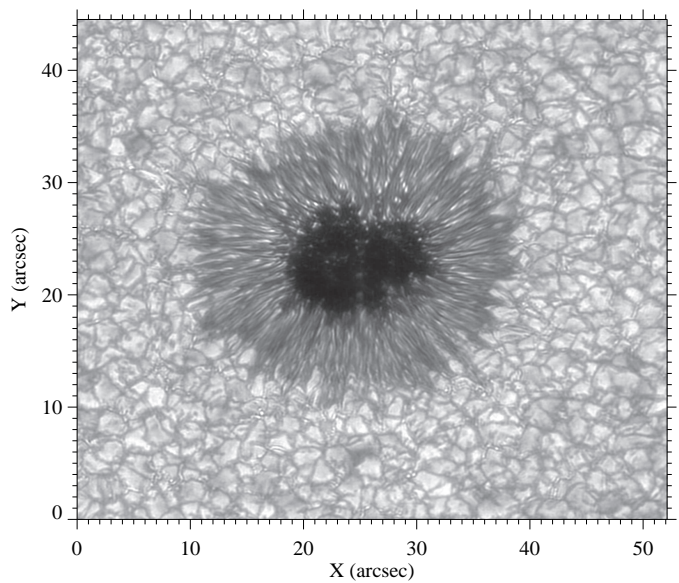}
  \caption{Speckle-restored blue continuum image at 4506 \AA\ of the sunspot in NOAA 12662 observed on 2017 June 17 at 09:13:33 UT with HiFI.}  
  \label{fig:HiFI}
\end{figure}

The target of the observations, the sunspot in active region NOAA 12662, was observed on 2017 June 17, when it was located at solar coordinates $x=-387\arcsec$, $y=179\arcsec$, with the 1.5-meter GREGOR solar telescope \citep{Schmidt+etal2012}. This is the same sunspot and date analyzed in \citet{Felipe+etal2018b}. Figure \ref{fig:HiFI} shows a contextual image of the sunspot acquired with the High-resolution Fast Imager \citep[HiFI;][]{Kuckein+etal2017} using a blue continuum filter (centered at 4506 \AA). The image restoration was performed with KISIP \citep{Woger+vonderLuhe2008}. In this work we focus on the spectroscopic images obtained with the GREGOR Fabry-P\'erot Interferometer \citep[GFPI;][and references therein]{Puschmann+etal2012} using $2\times2$ binning. We acquired intensity images at 20 wavelengths positions along the \FeI\ 5435 \AA\ line, with 8 accumulations per position and an exposure time of 20 ms each. The wavelengths are uniformly distributed along the profile of the line, with a spectral sampling of 26.2 m\AA. The full data series consist of 184 scans obtained from 08:10 to 09:20 UT, with a temporal cadence of about 23 s. The size of the field-of-view (FOV) is $55\arcsec.7\times41\arcsec.5$, and the image scale is 0\arcsec.081 pixel$^{-1}$. The data reduction was performed with sTools \citep{Kuckein+etal2017}, and the images were restored using Multi-Object Multi-Frame Blind Deconvolution \citep[MOMFBD][]{Lofdahl2002, vanNoort+etal2005}.

The \FeI\ 5435 \AA\ line is a magnetically insensitive line (effective Land\'e factor $\bar{g}=0$) formed at the high photosphere. \citet{BelloGonzalez+etal2010b} derived a formation height of 500 km in granules and 620 km in intergranules using NLTE modeling in numerical simulations of convection. From the analysis of phase differences between the velocity measured from different pairs of lines and assuming upward propagation of slow magnetoacoustic waves, \citet{Felipe+etal2018b} estimated a formation height of 542 km, while \citet{Chae+etal2017} obtained 280 km. In the following, we will focus on the Doppler velocity measured from a second order polynomial fit to the intensity core of the \FeI\ 5435 \AA\ line.  

A spectropolarimetric raster map of the sunspot was acquired with the GREGOR Infrared Spectrograph \citep[GRIS][]{Collados+etal2012} between 10:29 and 10:49 UT. The data correspond to the four Stokes parameters in the spectral region around 10830 \AA. The lines of interest are the \SiI\ 10827 \AA\ and the \CaI\ 10839 \AA\ lines. The pixel size along the slit is 0\arcsec.135, and the raster map was constructed from 200 consecutive slit positions shifted in increments of 0\arcsec.135 in the direction perpendicular to the slit. The dark and flat-field corrections and the polarimetric calibration were performed following the standard procedures \citep{Collados1999, Collados2003}. The GREGOR polarimetric calibration unit \citep{Hofmann+etal2012} was employed to obtain the data for the polarimetric calibration. 

The \SiI\ 10827 \AA\ and \CaI\ 10839 \AA\ lines were inverted independently using the Stokes Inversion based on Response functions (SIR) code \citep{RuizCobo+delToroIniesta1992}. In both inversions we have employed 5 nodes in temperature, 2 nodes in velocity and magnetic field strength, and 1 node in the azimuth and inclination of the magnetic field. The azimuth and inclination were chosen to be independent of the atmospheric height. We are mainly interested in the magnetic field geometry and its variation with height, from the deep photosphere (measured with the \CaI\ 10839 \AA\ line) to the mid-photosphere (measured with the \SiI\ 10827 \AA\ line). The results of the inversion have been transformed to the solar local reference frame.

\section{Results}
\label{sect:results}

\subsection{Evolution of a two-armed spiral}
\label{sect:evolution}

\begin{figure*}[!ht] 
 \centering
 \includegraphics[width=17cm]{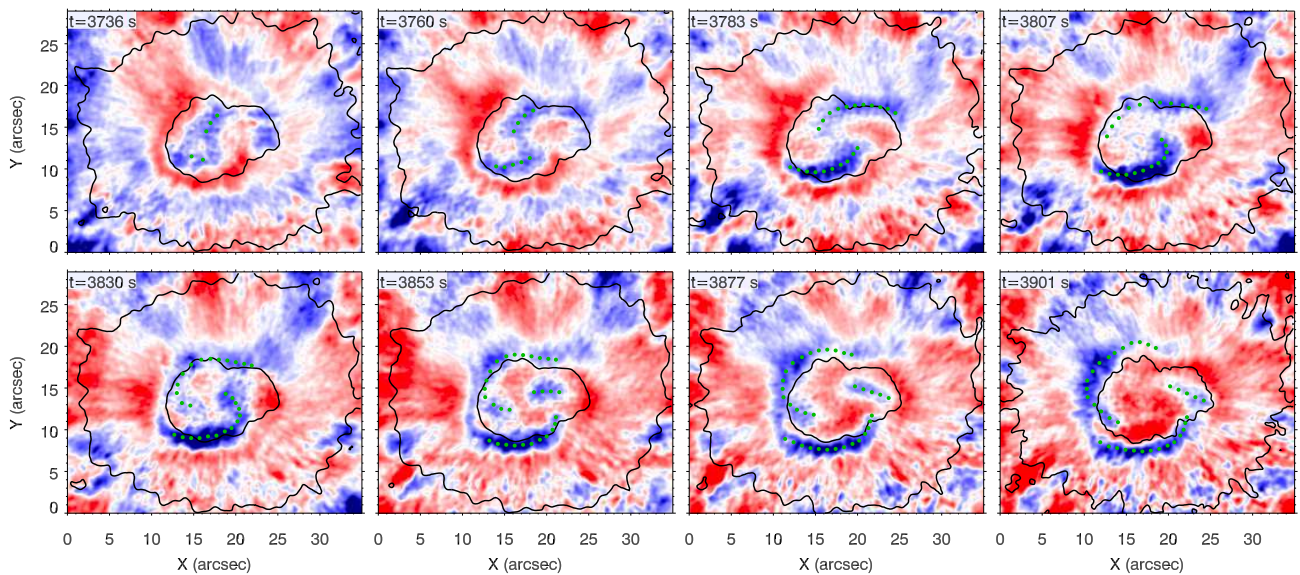}
  \caption{Temporal evolution of the Doppler velocity, inferred from the GFPI \FeI\ 5435 \AA\ restored images, during the appearance of a two-armed spiral wavefront. Blue (red) color indicates upflows (downflows). Velocities are saturated for amplitudes higher than 300 m s$^{-1}$. The black lines mark the umbral and penumbral boundaries, as determined from contours of constant intensity. The green circles highlight the location of the spiral wavefronts. The time at the top left corner of each panel shows the seconds since the beginning of the observations (08:10:08 UT). See movie in the online material.}  
  \label{fig:spiral_evolution}
\end{figure*}

At 09:12:16 UT ($t=3736$ s, as referred to the initial time of the data series), a two-armed spiral pattern starts to develop inside the umbra of the sunspot. This spiral structure is clearly visible in the Doppler velocity and its evolution is illustrated in Fig. \ref{fig:spiral_evolution} and in the online movie. This data is filtered in frequency, keeping waves with frequency between 2 and 8 mHz. The first panel of the figure shows that most of the umbra is filled with a slightly blueshifted atmosphere. During the next time steps, part of the blue wavefront at the bottom region of the umbra moves towards the right hand side, and the wavefront at the top part of the umbra moves towards the left hand side. For the sake of clarity, green circles are overplotted to track the wavefronts and enhance the sape of the spiral arms. The position of the green circles was defined manually by roughly selecting the middle of the blushifted wavefront. They move counterclockwise and towards the center of the umbra. Simultaneously, the tail of the spiral propagates radially outward. At $t=3807$ s the two spiral arms are defined and their origins are close to each other near the center of the sunspot.  

At $t=3830$ s the wavefronts of the spiral keep moving counterclockwise around the central part of the umbra, while the wavefronts near the umbra-penumbra boundary move radially outward, similar to running penumbral waves. As a result, an S-shaped wavefront is found at $t=3853$ s. In the last two panels illustrated in Fig. \ref{fig:spiral_evolution}, all the wave that once formed the blue wavefronts of the spiral are expanding in the radial direction. The red wavefronts in the last panel exhibit a spiral-like shape, similar to that shown by the blue wavefronts at $t=3807$ s, although it is less clear.

\subsection{Frequency filtered wavefronts}
\label{sect:filter}

\begin{figure*}[!ht] 
 \centering
 \includegraphics[width=17cm]{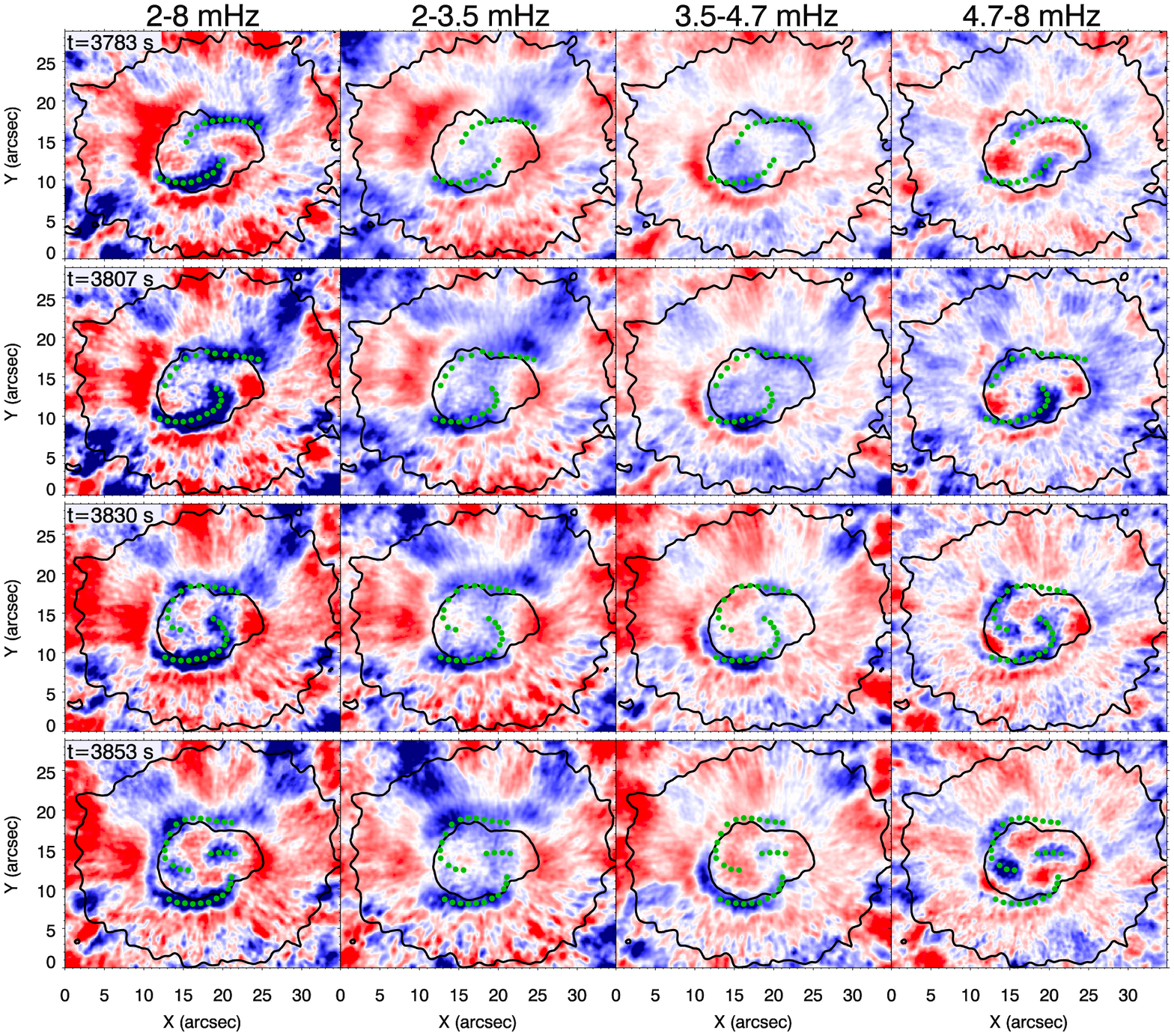}
  \caption{Temporal evolution of the Doppler velocity, inferred from the GFPI \FeI\ 5435 \AA\ restored images, during the appearance of a two-armed spiral wavefront, filtered in frequency bands. From left to right columns: 2-8 mHz, 2-3.5 mHz, 3.5-4.7 mHz, 4.7-8 mHz. The black lines and green circles have the same meaning from Fig. \ref{fig:spiral_evolution}.}      
  \label{fig:spiral_filtered}
\end{figure*}

Figure \ref{fig:spiral_filtered} shows four of the time steps (rows) when the spiral structure is visible filtered in different frequency bands (columns). The first column illustrates the Doppler shift after applying a broad frequency filter, including frequencies between 2 and 8 mHz (the same filter applied to data from Fig. \ref{fig:spiral_evolution}). The wavefronts of the different sections that compose the spiral are distributed among the frequency bands shown between second and fourth columns. This way, the outer parts of the spiral, which appears at \hbox{$t=3783$ s} at the umbra-penumbra boundary and then propagate towards the penumbra, corresponds mainly to waves in the low-frequency band (2-3.5 mHz), and partially to waves in the mid-frequency band (3.5-4.7 mHz). This shows that the low-frequency waves can propagate to the upper photosphere in regions of inclined magnetic field, since the cutoff frequency is reduced due to the lower gravity along the inclined magnetic field lines \citep{Bel+Leroy1977, Jefferies+etal2006}. On the contrary, inside the umbra the dominant part of the wavefront corresponds to the high-frequency waves (4.7-8 mHz). In regions with vertical magnetic field only waves above the cutoff can actually reach higher layers \citep[see][for an evaluation of the vertical variation of the cutoff frequency in the umbra of this sunspot]{Felipe+etal2018b}. The counterclockwise movement of the wavefronts and their displacement towards the center of the umbra is produced by these high-frequency waves.

This analysis reveals that the characteristic pattern of the two-armed spiral wavefront is composed by the contribution of waves with different frequencies. The inner part of the spiral, near the central part of the umbra, corresponds to high-frequency waves, while the outer parts are low-frequency waves. We find coherence between the spatial distribution of waves at different frequencies, in the sense that in regions of the wavefront where the wave amplitude in a certain band start to decrease (for example, the high-frequency waves near the umbra-penumbra boundary) the amplitude of the following band becomes stronger. This suggest that these waves are related, they propagate along neighbor magnetic field lines and probably share a common driving event. The atmosphere of the sunspots acts as a filter, allowing the propagation of certain frequencies at each position according to the magnetic field inclination, and shaping the wavefronts as the upward propagating slow magnetoacoustic waves move along the field lines. The apparent horizontal movement of the spiral wavefronts is produced by the relative delay of these waves in reaching the formation height of the \FeI\ 5434 \AA\ line. The spiral wavefronts sweep across this atmospheric layer.

\subsection{Magnetic field structure}
\label{sect:magnetic_field}

The magnetic field structure of the sunspot was derived from the inversions of the spectropolarimetric raster map obtained with GRIS after the GFPI temporal series was completed. The spiral structure discussed in the previous sections appeared at 09:12:16 UT, while the GRIS map was scanned between 10:29:24 UT and 10:49:00 UT. Although they are not simultaneous, we do not expect significant changes in the inferred magnetic field topology and that present during the spiral event.    

\begin{figure*}[!ht] 
 \centering
 \includegraphics[width=18cm]{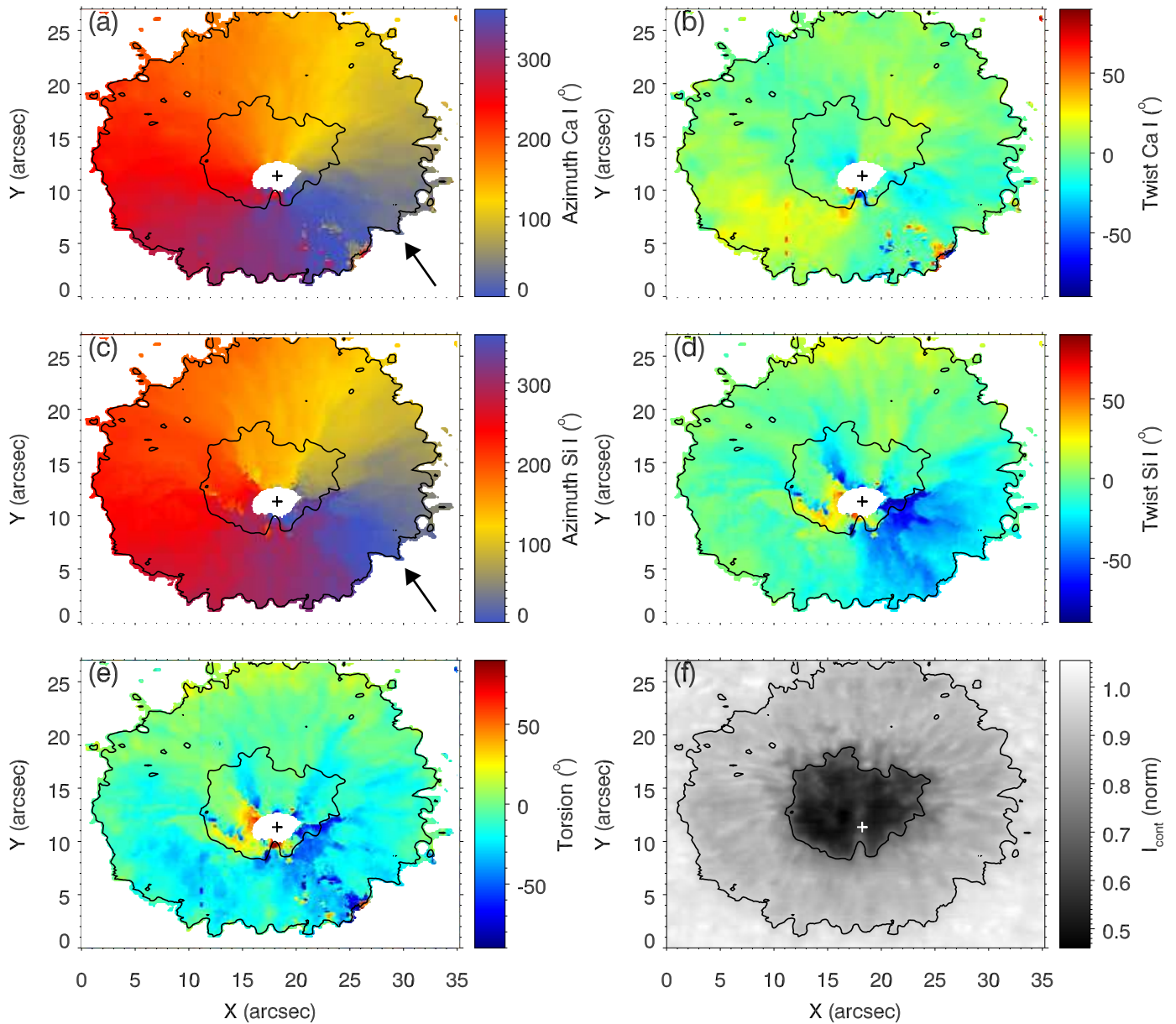}
  \caption{Physical parameters inferred from the GRIS observations. Top row: magnetic field azimuth (panel a) and magnetic twist (panel b) measured with the \CaI\ 10839 \AA\ line. Middle row: magnetic field azimuth (panel c) and magnetic twist (panel d) measured with the \SiI\ 10827 \AA\ line. Bottom row: magnetic torsion, defined as the difference between the azimuths obtained from the \SiI\ 10827 \AA\ and \CaI\ 10839 \AA\ lines (panel e), and continuum intensity from GRIS raster map (panel f). The azimuth is defined as the angle between the horizontal component of the magnetic field and the solar west-east direction. The arrow at the bottom right of panels a and c points to the direction of an azimuth value of $0^{\circ}$. The azimuth increases counterclockwise. Black lines indicate the umbral and penumbral boundaries, as determined from contours of constant intensity. The cross shows the position selected as the center of the sunspot for the calculation of the magnetic twist in panels b and d. In panels a-e, regions with quiet Sun intensity and the center of the spot, where the estimation of the azimuth is not reliable, have been masked out.}  
  \label{fig:azimuth_twist}
\end{figure*}

In this section we focus on the analysis of the azimuthal component of the sunspot magnetic field at two atmospheric heights. The inversions of the \CaI\ 10839 \AA\ line provide the magnetic and thermodynamic properties at the deep photosphere, while the \SiI\ 10827 \AA\ line is formed at higher photospheric layers \citep[see Fig. 4 from][for a representation of the response functions of these lines]{Felipe+etal2016b}. The azimuth at these two atmospheric heights is plotted in panels (a) and (c) from Fig. \ref{fig:azimuth_twist}. As expected for a sunspot with negative polarity, the horizontal magnetic field approximately points towards the center of the spot at both heights. There are some deviations from a radial configuration. Panels (b) and (d) from Fig. \ref{fig:azimuth_twist} show the magnetic twist measured from the \CaI\ 10839 \AA\ and \SiI\ 10827 \AA\ lines, respectively. We define the magnetic twist as the angle between the direction of the horizontal magnetic field at a certain location and the vector that links that location with the center of the sunspot. Note that in a sunspot like the present one it is not trivial to define a sunspot center. The umbra exhibits fine structure, including many umbral dots at the top part and a light bridge formed by several bright points (Fig. \ref{fig:HiFI}). We have chosen the center of the sunspot where the magnetic field, inferred from the \CaI\ 10839 \AA\ line, is vertical (indicated by a cross in Fig. \ref{fig:azimuth_twist}). Since the twist is not unequivocally defined, following \cite{SocasNavarro2005a} we have also defined the torsion as the vertical variation of the azimuth. Figure \ref{fig:azimuth_twist}e shows the torsion computed by subtracting the azimuth measured with the \CaI\ 10839 \AA\ line from the azimuth measured with the \SiI\ 10827 \AA\ line.

The magnetic field configuration at the deep photosphere shows a low twist in most of the umbra (Fig. \ref{fig:azimuth_twist}b). Only some penumbral regions exhibit a certain twist, which is generally below $30^{\circ}$. On the contrary, the twist measured at the height of the \SiI\ 10827 \AA\ line shows high values at the bottom right quarter, which reach values up to $-90^{\circ}$ around the umbra-penumbra boundary. This twist contrasts with that found inside the umbra at the bottom left quarter, which is mainly positive. \citet{SocasNavarro2005a} found that the magnetic twist can change the sign from the photosphere to the chromosphere. Our observations indicate that the twist can change from the deep photosphere (where it is almost negligible) to higher photospheric layers (where negative and positive twists coexist). This variation takes place over a height difference around 130 km \citep{Felipe+etal2018b}. However, the degree of the twist may be sensitive to the choice of the center of the sunspot. In this context, it can be more meaningful to discuss the magnetic field torsion. Figure \ref{fig:azimuth_twist}e shows that the bottom part of the penumbra has a slightly negative torsion, with stronger values near the umbra-penumbra boundary (in the same region where the twist from the \SiI\ 10827 \AA\ line was higher). Since the magnetic field at the formation height of the \CaI\ 10839 \AA\ line is approximately pointing towards the center, a negative torsion indicates that at the formation height of the \SiI\ 10827 \AA\ line the horizontal component of the magnetic field is shifted clockwise from the radial direction (see Fig. \ref{fig:Bhor_Vhor} for a vectorial representation of the horizontal magnetic field). A positive torsion, as that measured at the left and bottom part of the umbra, means that the magnetic field is shifted counterclockwise. The position on the horizontal plane of the wavefront of a slow magnetoacoustic wave propagating along magnetic field lines from the deep photosphere to the high photosphere will rotate according to the direction of the torsion. As discussed in Sect. \ref{sect:evolution}, the bottom arm of the spiral wavefront initially appears at the left side of the umbra, and it moves counterclockwise towards the center of the sunspot. This movement is in agreement with the positive torsion measured at that region from the analysis of the spectropolarimetric map. On the contrary, at the regions where the measured torsion is negative we do not detect clockwise motions of the spiral wavefront. In that part we only find a radially outward movement.

\subsection{Wavefront tracking}
\label{sect:tracking}

In order to evaluate the relation between the movement of the spiral wavefront and the direction of the magnetic field, we have measured the phase velocity of the wavefront in the horizontal plane. We have tracked the position of the wavefronts using a technique similar to local correlation tracking, which is a common approach for estimating the horizontal velocity field in the solar surface from the evolution of intensity images. The methodology is as follows. We focus on two Doppler velocity maps, measured at two consecutive time steps. For each spatial position in the first map, we have derived the horizontal velocity that best describe its position in the second map. This is done in three steps: First, we have isolated the neighborhood around the location of interest and we have applied a Gaussian mask to reduce the relevance of regions far away from that location in both Doppler maps. Second, we have computed the correlation between both maps sampling different displacements in the two horizontal directions. Third, we have located the peak of the cross correlation. Once the optimal displacement is found for each pixel, the horizontal velocity is derived by dividing it by the time difference between both maps. This process has been performed for all locations inside in the umbra and surroundings for all the maps when the spiral wavefronts are present. 

\begin{figure}[!ht] 
 \centering
 \includegraphics[width=9cm]{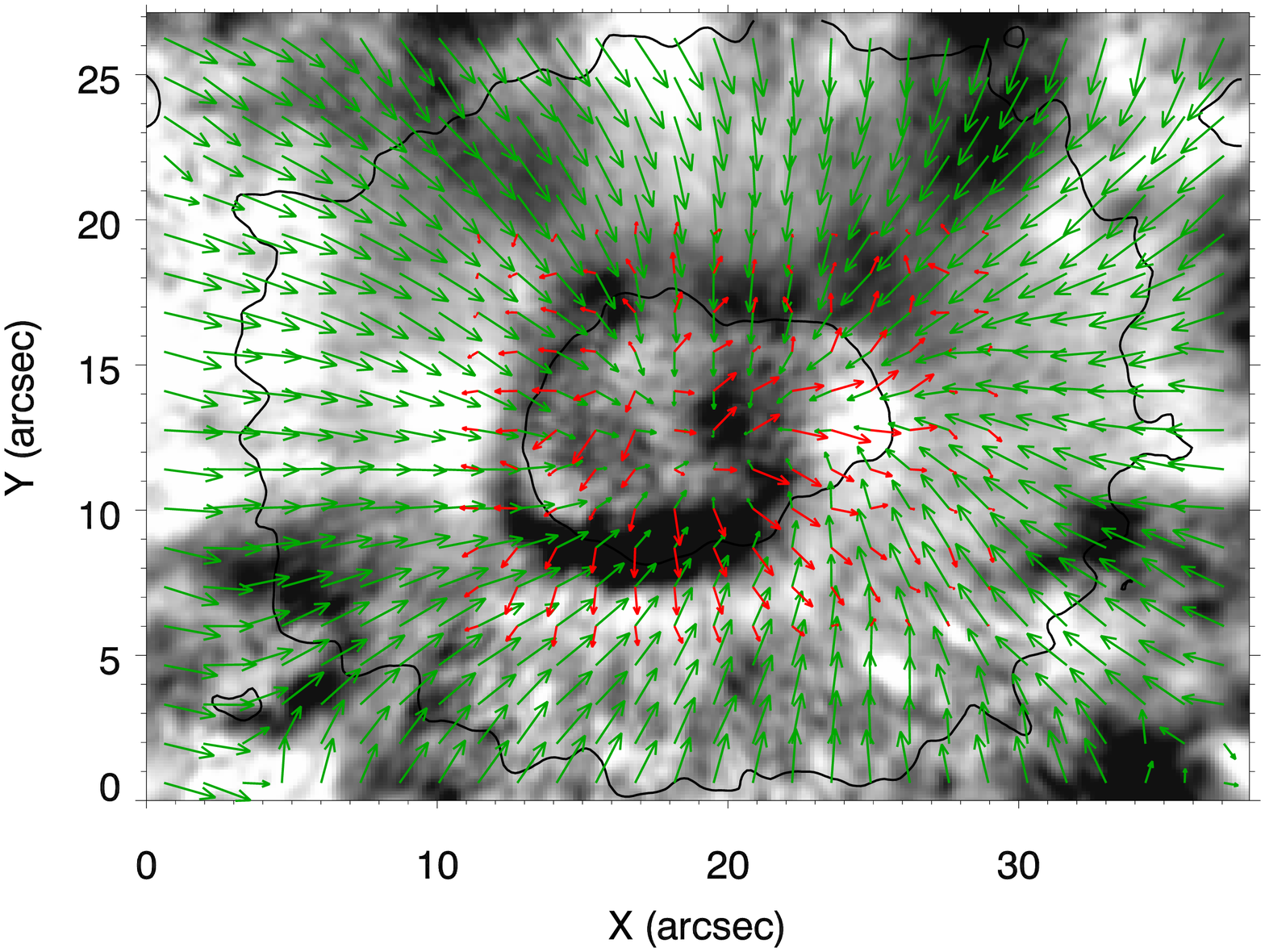}
  \caption{Doppler velocity map inferred from the GFPI \FeI\ 5435 \AA\ restored images at $t=3830$ s. Black color indicates upflows and white color corresponds to downflows. Green arrows display the direction and magnitude of the horizontal magnetic field inferred from GRIS \SiI\ 10827 \AA\ observations and red arrows show the direction and magnitude of the wavefronts phase speed determined by tracking the wavefronts from GFPI data. The longest red arrows correspond to a phase velocity of 20 km s$^{-1}$.}      
  \label{fig:Bhor_Vhor}
\end{figure}

Figure \ref{fig:Bhor_Vhor} shows a map of the vectorial horizontal velocity field obtained as the median of the vector fields derived from the eight time steps illustrated in Fig. \ref{fig:spiral_evolution} (red arrows). This field is overplotted on top of one of the Doppler velocity maps when the spiral wavefronts are visible, and together with the vectorial horizontal magnetic field obtained from GRIS \SiI\ 10827 \AA\ data (green arrows). Cospatial maps from data obtained with GRIS (magnetic field structure) and data retrieved with GFPI (LOS velocity field) have been constructed by rescaling GRIS continuum intensity to the same spatial scale of GFPI data, and looking for the spatial displacement that provides the best correlation between this map and the continuum intensity obtained with GFPI during the raster scan of GRIS data. Once this displacement is derived from intensity images, the same transformation has been applied to the magnetic field maps.

\begin{figure}[!t] 
 \centering
 \includegraphics[width=9cm]{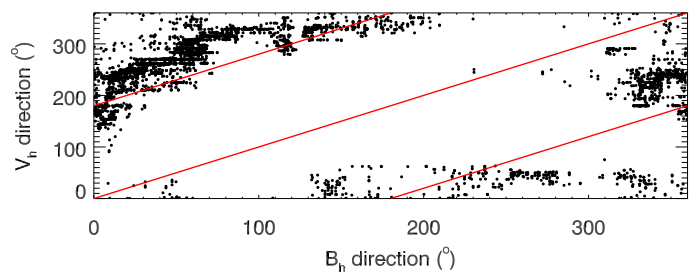}
  \caption{Angle of direction of the horizontal velocity of the spiral wavefronts as a function of the direction of the horizontal magnetic field. The angle is measured counterclockwise from the $X$ axis in Fig. \ref{fig:Bhor_Vhor}. Red lines indicate the regions where the wavefront velocity is aligned with the horizontal magnetic field. The middle line represents the case where the velocity and the magnetic field point to the same direction, while the two outer lines correspond to vectors in opposite directions.}      
  \label{fig:angle_Bhor_Vhor}
\end{figure}

The horizontal velocity of the spiral wavefronts follows approximately the direction of the magnetic field at most spatial positions, but pointing in opposite directions. This is expected, since this sunspot has a negative polarity. The magnetic field lines are pointing radially inward and downward, but the slow magnetoacoustic waves propagate upward along those field lines. The horizontal wavefronts at a fixed height are found to perform an approximately radially outward movement, since they are the visual pattern of the upward propagating wave that reach the formation height of the \FeI\ 5434 \AA\ line at different times. There are systematic departures of the horizontal velocity from a perfect alignment with the horizontal magnetic field. For a better visualization, Fig. \ref{fig:angle_Bhor_Vhor} shows a scatter plot between the angles characterizing the direction of both vector fields. It exhibits a strong correlation between the direction of the velocity and the horizontal magnetic field. Almost all the points concentrate around the lines that mark the position where both vectors are aligned but pointing in opposite directions. There is a preference for values of the velocity angle higher than that representing a perfect alignment. This means that the velocity direction is shifted counterclockwise from the direction of the magnetic field, as seen in most locations from Fig. \ref{fig:Bhor_Vhor}. We point out two causes for this displacement. First, the magnetic field is determined at the formation height of the \SiI\ 10827 \AA\ line, while the velocity field is derived from the \FeI\ 5434 \AA\ line. \citet{Felipe+etal2018b} found a difference of 200 km in the formation height of both lines inside a sunspot umbra from the analysis of wave propagation between both layers. As shown in Fig. \ref{fig:azimuth_twist}, the twist of this sunspot change between the deep photosphere and the photosphere, and it is expected to be different at the upper photosphere where the velocity is measured. Second, and more important, we have not measured the real horizontal velocity of the waves as they propagate along magnetic field lines. We have tracked the apparent movement of the wavefronts at a certain height. Part of this horizontal movement is caused by the delay between different waves that propagate upward along neighbor field lines. This is not necessary aligned with the orientation of the horizontal magnetic field.

\section{Discussion and conclusions}
\label{sect:conclusions}

Several works have reported the presence of spiral-shaped wavefronts in sunspots \citep{Sych+Nakariakov2014,Su+etal2016}. Although they agree to identify these waves as upward propagating slow magnetoacoustic waves, there is not a clear picture describing the spirality of the wavefronts. \citet{Sych+Nakariakov2014} concluded that this phenomenon was not associated to the magnetic twist, based on the absence of twist in the magnetic geometry of the sunspot that they studied \citep{Reznikova+Shibasaki2012}. On the contrary, \citet{Su+etal2016} identified different spiral structures, some of them with one arm and other with two or three arms. They proposed that the one-armed spiral may be caused by the reflections of the wavefronts at a light bridge, and the multi-armed wavefronts may be produced by the twist of the umbral magnetic field beneath the photosphere.

In this paper, we have identified a new event of spiral wavefronts in a sunspot umbra. We have focused on the analysis of oscillations in Doppler velocity, opposite to the previous works which have mainly studied intensity oscillations. The wavefronts of interest initially appear as a blueshifted region inside the umbra, and they expand in the counterclockwise direction, forming a two-armed spiral with its origins near the sunspot center. The spiral also exhibits outward propagation, and became running penumbral waves around the umbral boundary. Thus, they propagate radially outwards across the penumbra.

In order to evaluate the impact of the magnetic field geometry on the shape of the wavefronts, we have derived it from multi-height spectropolarimetric inversions of the sunspot. Our results show that the magnetic twist is negligible at the deep photosphere, whereas it is significant at higher photospheric layers. The twist not only changes with height, but also presents positive and negative values at the same layer. We have proven that the motions of the spiral wavefronts are approximately aligned with the direction of the horizontal magnetic field. This way, the presence of magnetic twist is relevant for determining the shape of the observed wavefront. Interestingly, the sign of the twist in the bottom left part of the umbra agrees with the counterclockwise movement of the spiral wavefront in that region. However, this is not the case in other parts of the umbra where the spiral arms display a similar behavior. Although the twist must have a significant impact on the shape of the observed wavefronts, we cannot conclude that it is the main cause of the spiral structure of the oscillations. 

The analysis of the spiral wavefronts filtered in different frequency bands reveals that the whole structure that we have observed is composed by the contribution from waves with different frequencies. Higher frequencies dominate inside the umbra, whereas the amplitude of lower frequencies become more important at regions farther away from the center of the sunspot. The shape and apparent motion of the wavefronts are determined from the coherence between slow magnetoacoustic waves that propagate upward along different field lines. These waves are filtered due to the gravitational stratification of the atmosphere and the presence of a cutoff, and this filtering is modulated by the inclination of the magnetic field, which reduces the cutoff value as the effective gravity along field lines decreases. 
All these waves form the spiral pattern as they coherently reach the observed atmospheric height, with the apparent motions determined by their relative delays. We speculate that the observed pattern may be a caused by the fine structure of the sunspot, which exhibits regions with many umbral dots and a light bridge (Fig. \ref{fig:HiFI}). The local variations of the magnetic field geometry associated to these features may be relevant to deform the field lines and produce spatial variations in the cutoff frequency, which can shape the wavefronts.


\begin{acknowledgements} 
Financial support from the Spanish Ministry of Economy and Competitivity through projects AYA2014-55078-P, AYA2014-60476-P and AYA2014-60833-P is gratefully acknowledged. C.K. was supported in part by grant DE 787/5-1 of the German Science Foundation (DFG). The 1.5-meter GREGOR solar telescope was built by a German consortium under the leadership of the Kiepenheuer-Institut f\"ur Sonnenphysik in Freiburg with the Leibniz-Institut f\"ur Astrophysik Potsdam, the Institut f\"ur Astrophysik
G\"ottingen, and the Max-Planck-Institut f\"ur Sonnensystemforschung in G\"ottingen as
partners, and with contributions by the Instituto de Astrof\'isica de Canarias and
the Astronomical Institute of the Academy of Sciences of the Czech Republic.
\end{acknowledgements}

\bibliographystyle{aa} 
\bibliography{biblio.bib}

\end{document}